# 3D Nanomagnetism in Low Density Interconnected Nanowire Networks

Edward C. Burks,[1] Dustin A. Gilbert,[1, 2,3] Peyton D. Murray,[1] Chad Flores,[1] Thomas E. Felter,[4] Supakit Charnvanichborikarn,[5] Sergei O. Kucheyev,[5] Jeffrey D. Colvin,[5] Gen Yin,[6] and Kai Liu[1,6,*]

[1]*Physics Department, University of California, Davis, CA 95618*

[2]*Department of Materials Science and Engineering, University of Tennessee, Knoxville, TN 37996*

[3]*Department of Physics and Astronomy, University of Tennessee, Knoxville, TN 37996*

[4]*Sandia National Laboratories, Livermore, CA 94551*

[5]*Lawrence Livermore National Laboratory, Livermore, CA 94551*

[6]*Physics Department, Georgetown University, Washington, DC 20057*

**Abstract**

Free-standing, interconnected metallic nanowire networks with density as low as 40 mg/cm$^3$ have been achieved over cm-scale areas, using electrodeposition into polycarbonate membranes that have been ion-tracked at multiple angles. Networks of interconnected magnetic nanowires further provide an exciting platform to explore 3-dimensional nanomagnetism, where their structure, topology and frustration may be used as additional degrees of freedom to tailor the materials properties. New magnetization reversal mechanisms in cobalt networks are captured by the first-order reversal curve method, which demonstrate the evolution from strong demagnetizing dipolar interactions to intersections-mediated domain wall pinning and propagation, and eventually to shape-anisotropy dominated magnetization reversal. These findings open up new possibilities for 3-dimensional integrated magnetic devices for memory, complex computation, and neuromorphics.

**Introduction**

Porous metallic nanostructures have exciting potential applications in such areas as lightweight materials,[1-2] energy conversion and storage,[3-6] and filters.[7-8] Networks of interconnected magnetic nanowires could provide an exciting platform to explore 3-dimensional (3D) nanomagnetism, where their structure, topology and frustration may be used as additional degrees of freedom to tailor the materials properties.[9-12] For example, cylindrical nanowire building blocks present opportunities for exploring curvature-driven Dzyaloshinskii–Moriya interaction (DMI),[13] exotic types of 3D topological spin textures[9, 14-16] and chiral symmetry breaking;[17-19] such networks could be model systems to study 3D "racetrack" memories,[20-21] where the information is encoded in DWs propagating along the nanowires, and the intersections may be used to manipulate DWs, such as pinning, annihilation or changing their topological character. Recent studies have shown the possibility of addressing DWs in a contactless fashion via chemisorption,[22] which is attractive for 3D systems. These networks may also form fascinating platforms for studying artificial spin ice,[23-27] where the interplay between dipolar interaction among segments of the nanowires and the interconnected nanowire geometry gives rise to magnetic frustration and the existence of a large number of magnetic configurations, with potential applications in complex computation.[28] Such networks also can be used to modify electric and thermal transport properties, with potentials for spin caloritronics applications.[29-31] Furthermore, the connectivity and complexity of the networks would present interesting opportunities to explore artificial neural networks, e.g., the resistive switching properties of the intersections of the network may be used to activate / deactivate the network.[32]

In this work we report the study of 3D nanomagnetism in quasi-ordered metallic networks realized over cm-scale areas using electrodeposition into ion-track-etched nanoporous membranes, and feasibility demonstration of 3D information storage. Using a cost-effective multiple azimuthal angle ion-tracking and etching method with suitable fluence, interconnected nanoporous polycarbonate membranes were first fabricated. A subsequent electrochemical deposition process led to interconnected, quasi-ordered metallic nanowire networks, with density as low as 40 mg/cm$^3$. The unique structure of the networks provides a fascinating platform for explorations of 3D nanomagnetism and spin textures. The interplay amongst shape anisotropy, dipolar interactions and domain wall pinning and propagation through interconnections leads to the interesting possibility of accessing multiple magnetic states in such 3D assemblies by varying the applied field. These magnetic states may be further used to encode information and be driven to propagate through the network, opening up new opportunities for 3D magnetic memory and logic, probabilistic computing and neuromorphics.

**Template production**

Circular polycarbonate membranes with thicknesses of 3 μm and 6 μm were first irradiated over ~ 1 inch$^2$ areas with 19.6 MeV Xe$^{6+}$ ions at multiple angles with equal fluence, including normal incidence (tracking direction 1) and up to four 45° incidence angles (the colatitude angle) at symmetric azimuthal directions (0°, 90°, 180°, and 270°), as illustrated in **Figure 1a**.[33-34] The total ion irradiation fluence was varied from 5×10$^7$ ions/cm$^2$ to 4×10$^9$ ions/cm$^2$. The energetic particles penetrate through the entire thickness of the membranes and create latent tracks of ion damage (**Figure 1b**). Next, each side of the membrane is exposed to an ultraviolet (UV) ozone cleaning system for 3 minutes. This step further breaks the crosslinking in the damaged areas, making these regions more susceptible to chemical etching.[35] The tracks were then etched with intermittent sonication in 6 M NaOH. The resultant voids form an intersecting nanopore network (**Figure 1c**). In this work, nanoporous templates were created with pore diameters ranging from 75 to 350 nm.

**Synthesis of interconnected nanowire networks**

The interconnected nanowire foam was fabricated by electrodeposition to fill the nanoporous network. A Cu layer (up to 500nm) was first magnetron sputtered onto one side of the membrane to serve as the working electrode (**Figure 1d**). Electrodeposition of Cu and Co nanowires was performed using known electrolyte recipes and procedures[36-38] (**Figure 1e**). In order to create a free-standing metallic structure, the polycarbonate membrane was etched by submerging it in dichloromethane for a few seconds; to then transfer the light weight network structure into air while retaining its volume, a freeze drying process was used to prevent structural collapse due to surface tension.[6, 39] A liquid exchange process was performed to transfer the sample to water, which was subsequently frozen with liquid nitrogen. The frozen sample was placed in rough vacuum ($\approx$10 mTorr) to sublimate the ice away, leaving the Cu or Co structure intact and in air (**Figure 1f**).

The morphology of the nanoporous membranes and the resultant nanowire networks was examined by scanning electron microscopy (SEM) using a Hitachi 4100 SEM. The interconnecting nanowire networks consist of nanowire arrays with well-defined geometrical alignment on top of the bottom electrode, as shown in **Figure 2**. A zoomed-in view of one of the intersections is shown in **Figure 2a inset**, illustrating that the nanowires have indeed grown into the irradiation-created tracks. The resultant network structure was held together by the thin working electrode and wire intersections ($\approx 10^9$ intersections/cm$^2$, see Supporting Information). X-ray diffraction measurements (not shown) indicate that the electrodeposited nanowires are polycrystalline, similar to earlier reports on similarly prepared nanowire arrays.[36-37]

The density of each sheet of the interconnected networks, including the bottom electrode, typically ranges from 100 mg/cm$^3$ to 500 mg/cm$^3$. To test the density limit, the grown networks embedded in the polycarbonate membrane with a 50 nm Cu electrode were folded onto itself and placed in dichloromethane to remove the polycarbonate, followed by the freeze-drying process. This leads to metallic Cu and Co nanowire networks, consisting of multiple sheets of the free-standing structure, with density as low as 40 mg/cm$^3$ (99% porosity) over a 0.1 cm$^3$ volume.

**Magnetization reversal in Cobalt networks**

Magnetic properties of the Co networks were measured by vibrating sample magnetometry (VSM) at room temperature. The external field, up to 1.5 Tesla, is applied at an angle $\theta$ relative to the membrane plane. For comparison, both intersecting nanowire networks and non-intersecting, parallel nanowire samples were investigated, which utilized Co nanowires with 75 nm diameter. The parallel nanowire samples were only ion-tracked normal to the plane of the membrane, thus all wires were parallel to one another. For the intersecting Co nanowire networks, the ion-tracking was performed at three angles: one normal to the membrane, and two at 45° colatitude angle from the membrane normal, at 0° and 180° azimuthal orientations, respectively (Figure 3 inset); all angles were exposed to the same irradiation fluence. The nanowires are much more likely to intersect with one another within the irradiation plane, as illustrated in the Supporting Information (**Figure S1**). The resultant intersecting nanowire lattices are shown in **Figure 2d**.

Details of the magnetization reversal were investigated using the first-order reversal curve (FORC) method.[40-46] In this measurement, the sample is first positively saturated, then brought to a reversal field, $H_R$, along the descending branch of the major hysteresis loop. Subsequently the applied field, $H$, is increased from $H_R$ to positive saturation, capturing the evolution of the magnetization, $M$, along a FORC. This process is repeated for a series of $H_R$ and a family of FORCs is measured. Details of the magnetization reversal process are captured in the FORC distribution: $\rho(H, H_R) \equiv -\frac{1}{2}\frac{\partial^2 M}{\partial H \, \partial H_R}$. Since sweeping of $H_R$ to more negative values probes the down-switching events, and measurement along increasing $H$ probes up-switching events,[45] the FORC distribution can also be presented in terms of local coercivity, $H_c = (H-H_R)/2$, and bias, $H_b = (H+H_R)/2$. For this work, the FORC distribution will be presented in the ($H_c$, $H_b$) coordinate due to the natural symmetries of the FORC features.

Magnetometry measurement of the parallel nanowires at $\theta = 0$ (field perpendicular to the wires) exhibits a tilted, hard-axis loop with a small remanence (**Figure 3a**). For comparison, at $\theta = 90°$ (parallel to wires), a square hysteresis loop with near-unity remanence is observed; at $\theta = 45°$, the loop is in between that of $\theta = 0$ and 90°. These loops are consistent with the shape anisotropy of the non-intersecting nanowires, which favors magnetization alignment along the nanowires.[38]

For the intersecting wire samples, magnetic measurements are first taken as a function of the angle between the applied field and the ion-tracking plane normal, $\varphi$, as shown in **Figure 3b** and **3d** inset. At $\varphi = 0$, i.e., field perpendicular to the plane of ion tracking, a hard axis loop is observed. One might expect the loop to be the same as that of the $\theta = 0$ loop for the parallel nanowires shown in Figure 3a, as the magnetic field is orthogonal to all the wires in both cases. However, while similar hard axis loops are observed, the coercivities are significantly different, 270 Oe for the parallel nanowires vs. 570 Oe for the intersecting wires, demonstrating a clear difference in the magnetization reversal behavior. This contrast highlights the effect of the intersections on the magnetic properties. At $\varphi = 45°$, a loop with much higher remanence is observed. At $\varphi = 90°$ (also $\theta = 90°$), i.e., field in the plane of ion tracking, a square hysteresis loop is observed, indicating an easy axis direction. The coercivity is 600 Oe, much larger than the 460 Oe in the parallel nanowires measured at $\theta = 90°$.

While the field stays in the irradiation plane ($\varphi = 90°$), as $\theta$ is reduced from 90°, the loops becomes more slanted and the remanence reduces at $\theta = 45°$ and 0° (**Figure 3c**). This evolution is a manifestation of the network shape anisotropy as well as interwire coupling. Specifically, at $\theta = 0°$, 1/3 of the wires are perpendicular to the field and 2/3 are at 45° from the applied field; at $\theta = 45°$, 1/3 of the wires are perpendicular, 1/3 are parallel, and 1/3 are 45° from the applied field; at $\theta = 90°$, 1/3 of the wires are parallel, and 2/3 are 45° from the applied field. As one might expect, the orientation with the largest nanowire projection along the magnetic field ($\theta = 90°$) remains the magnetic easy axis, as indicated by the large remanence. During the rotation, the shape anisotropy would typically dominate the reversal behavior of the wires; however, the intersections allow domains to traverse from wires more-aligned with the magnetic field into those less-aligned. Also, rotating the sample in the plane of the irradiation causes little change to the coercivity, in contrast to those shown in Figure 3a or 3b.

These magnetization reversal differences between parallel nanowires and the intersecting wires are more significant than possible contributions due to sample-to-sample variations, or a mere superposition of different measurement geometries. The latter effect is illustrated by using the parallel nanowire results shown in Figure 3a to attempt and reconstruct the intersecting case shown in Figure 3c. For example, an attempt to reconstruct the $\theta = 0$ loop in Figure 3c is made by summing 1/3 of the $\theta = 0$ and 2/3 of the 45° loops in Figure 3a for the parallel wires. The "reconstructed" loop and measured loop are shown in red and black, respectively, in Figure 3d. Clearly, the reconstruction fails to model the experimental measurements. This further emphasizes the importance of the network structure to the magnetization reversal process. Likely there are additional ferromagnetic coupling between the intersecting nanowires, which cause neighboring wires to reverse collectively. The numerous intersections may also act to impede magnetization reversal via domain wall movement, leading to the coercivity enhancement observed, similar to those seen in earlier 2-dimensional magnetic networks.[47]

**FORC analysis of Cobalt networks**

To further investigate details of the magnetization reversal process and effects of the intersections on the magnetic properties, the FORC method has been used.[40-43, 45-46, 48-49] For non-intersecting, parallel nanowires, the applied magnetic field was varied as a function of $\theta$, and the corresponding FORC distributions are shown in **Figure 4a-c**. At $\theta = 90°$, field parallel to the nanowires, a pair of elongated FORC ridges are observed along the $H_b$ axis (Figure 4a). The primary vertical ridge around $H_c = 0.7$ kOe is characteristic of demagnetizing dipolar interactions between nanowires.[50-51] Its maximal spread along the $+H_b$ direction quantitatively measures the maximal dipole field strength, which in this case is 1100 Oe.[45, 52] This ridge also has a small tail along the $H_c$ axis, indicative of intrinsic coercivity distribution due to variations in nanowire diameter, length, and grain size, as well as mixed crystalline phases and orientation;[52-53] the coercivity distribution also causes a slight tilt of the FORC ridge from an exact vertical alignment, and the difference in its spread in $+H_b$ and $-H_b$ directions.[45] This FORC ridge indicates that the nanowires reverse in isolated switching events, likely a domain nucleation then rapid propagation through the entire wire. In addition, there is a second, weaker FORC ridge at $H_c = 0$, indicating reversible behavior.[54-55] At $\theta = 45°$, qualitatively the same pair of FORC features are still present, indicating that the reversal still occurs as a single switching event, but the reversible FORC feature at $H_c = 0$ becomes much more significant (Figure 4b). Finally, at $\theta = 0°$, only the reversible FORC feature remains, indicating that the magnetic moments are being forced to orient orthogonal to the nanowires along this magnetic hard axis (Figure 4c).

For the intersecting wires, FORC measurements were first taken in the irradiation plane ($\varphi=90°$). At $\theta = 90°$ sample, again two FORC features are observed, similar to those in the parallel nanowire case, except that the vertical ridge at finite $H_c$ has evolved into a "wishbone" feature with a narrower spread along $H_b$ and a more pronounced tail along $H_c$ (**Figure 4d**). As noted above, the narrower distribution in $+H_b$, now spanning to 800 Oe, indicates a reduced dipolar interaction field – as would be expected from the new 45° segments. Note that the spread of the FORC feature along the $H_c$ axis is approximately equal to the intrinsic coercivity distribution plus the interaction field, and the latter is determined by the extent of the FORC ridge in $+H_b$.[45] Following this analysis, the intersecting wire sample possesses a similar intrinsic coercivity distribution as the non-intersecting wire sample (Figure 4a), and the appearance of a more pronounced tail along $H_c$ is a reflection of the weaker interactions that make the tail feature stand out more.

Rotating this sample in the irradiation plane, as $\theta$ is reduced to 45° (**Figure 4e**), the "wishbone" feature remains visible, but the spread along $H_b$ further shrinks, consistent with the expected reduction in the dipolar interactions. At $\theta = 0°$ (**Figure 4f**), the FORC feature collapses to a symmetric dome centered on the $H_c$ axis. This feature corresponds to the limit of no dipolar interactions, and the spread in $H_c$ identifies the intrinsic coercivity distribution.[45] This type of FORC distribution is often associated with domain wall propagation,[54] implying that the connections between wires are facilitating reversal across the planes of intersection in the sample. Interestingly, the comparison between the intersecting and parallel wires (Figure 4c) shows that the distribution never collapses to the reversible-only FORC feature at $H_c = 0$. This is due in-part to the oblique irradiation, which enforces that only 1/3 of the wires are orthogonal to the field, placing a limit on the reversible feature. We also compare the intensity of the reversible feature in Figure 4f, in which 2/3 of the wires are at 45° and 1/3 are at 90° relative to the field, to the corresponding isolated wires, Figure 4b and 4c, and find that the FORC features are not simply a sum of the constituent elements. This difference is illustrative of the collective magnetization reversal resultant from the intersection of the wires.

A similar series of measurements was performed at different $\varphi$ angles while $\theta$ is kept at 90°. At $\varphi = 90°$, the same geometry as that in Figure 4d, the FORC distribution is remeasured and included in **Figure 4g** for completeness. At $\varphi = 45°$ (**Figure 4h**), the wishbone structure persists, with slightly stronger reversible behavior, consistent with the increased orthogonal alignment to the wires. At $\varphi = 0°$ (**Figure 4i**), the field is perpendicular to the irradiation plane, again the FORC feature collapses to the reversible ridge at $H_c = 0$, reproducing the parallel wire case when measured with an orthogonal field along the magnetic hard axis (Figure 4c). In both cases, the shape anisotropy dominates the magnetic orientation, resulting in the reversible behavior captured by the FORC diagrams. The continuous shift in weight between the hysteretic FORC feature (at $H_c > 0$) and the reversible feature (at $H_c = 0$) can be seen in Figure 4g-i to directly follow Figure 4a-c, again, highlighting the similar underlying mechanics, namely the shape anisotropy of the wires. However, rotating the field relative to the irradiation plane (Figure 4g-i) shows a clear effect from the intersecting wires, as the demagnetizing dipolar field reduces.

**Discussions**

The FORC diagrams in Figure 4 capture vividly the sensitive interplay of dipolar interactions, shape anisotropy and interconnection-mediated domain wall propagation as a function of applied magnetic field strength and geometry, especially the 3D perspectives shown in **Figure 4j-4r**. Strong demagnetizing dipolar interactions are found when most of the nanowires are aligned with the magnetic field (Figure 4a). Such dipolar effect has been shown to achieve analog memory effects in arrays of parallel magnetic nanowires.[50] Shape anisotropy dominates when the field is perpendicular to all the wires, leading to reversible FORC ridge (at $H_c = 0$) in Figure 4c and 4i. The intersecting nanowire network suppresses the dipolar interactions, as certain fractions of the nanowires are forced to be misaligned with the field (Figure 4d); and eventually leads to domain wall propagation through the intersections when the degree of misalignment is the greatest (Figure 4f). *Thus by varying the applied magnetic field strength, orientation, and sequence, it is in principle possible to selectively address a certain subset of the nanowire networks.* For example, in the geometry shown in Figure 4d, a suitable field may be used to only align 1/3 of the nanowires that are parallel to the field into single domain state (while this field is insufficient to saturate the other 2/3 of the nanowires that are 45° misaligned); after a subsequent rotation of the network/magnetic field to the geometry depicted in Figure 4e, the magnetic field may be tuned to address the 1/3 newly aligned branches or drive domain walls from interconnections with the previous single domain nanowires into these branches; alternatively, if geometry Figure 4f is used, the domain walls may propagate in 2/3 of the 45° misaligned nanowires, and the resultant magnetic state can be distinguished from that in Figure 4e. Therefore, it might be possible to encode digital information into the magnetic state of the networks, and propagate it through the interconnected magnetic network for potential 3D magnetic memory and logic applications. Furthermore, the magnetic state would reflect the probability of a domain wall to propagate through certain set of interconnections, which could be utilized for probabilistic computing.

Such network of magnetic nanowires also has the potential to implement repeatable multi-state memristors in neuromorphic circuits. When a magnetic domain is established in an isolated nanowire, a longitudinal current can drive the domain wall motion if the torques exerted on the domain wall overcome the pinning energy, which is usually contributed by disorders, anisotropy and the local effective exchange field provided by the neighboring spins. Therefore, to maintain a domain wall motion, the driving current density should exceed a threshold value, $j_{TH}$. In a nanowire network with intersecting points, however, the domain wall driven by $j_{TH}$ can be trapped at the intersections. This is due to a larger effective exchange field experienced by the spins at the intersections. These spins have more neighbors provided by the intersecting wires, and are in principle more difficult to switch. To further drive the domain wall, the current density should exceed another threshold value, $j'_{TH} > j_{TH}$. Thus, a time sequence of current density $j(t)$ that contains a series of spikes can implement a step-by-step domain-wall motion through the network. Unlike the scenario in conventional race-track memories, in an intersecting network the domain wall motion is discrete. Such discrete configuration can remember the time sequence of the current density if the width and the height of each spike are engineered to a "sweet" spot, resulting in an ideal candidate for multi-state memristors. This functionality may enable synapse capability to the nanowire network as an artificial neural network.

In summary, nuclear track-etched membranes irradiated at multiple azimuthal angles were used as an electrodeposition template to fabricate interconnected low density metallic nanowire networks. Free-standing copper and cobalt networks were grown with densities as low as 40

mg/cm$^3$. Details of the magnetization reversal characteristics of interconnected cobalt nanowire networks are captured in FORC diagrams, which demonstrate the evolution from strong demagnetizing dipolar interactions to intersections-mediated domain wall pinning and propagation, and eventually to shape-anisotropy dominated magnetization reversal in different geometries. Our findings demonstrate a 3D model system for explorations of nanomagnetism and spin texture, with potential applications in 3D magnetic memories and logic devices, complex computing, and memristors for neuromorphics.

**Supporting Information**.

The Supporting Information is available free of charge.

Estimation of number of intersections per unit area (PDF).

**AUTHOR INFORMATION**

**Corresponding Author**

* Kai Liu - Email: Kai.Liu@georgetown.edu

**ACKNOWLEDGEMENT**

This work has been supported by the NSF (ECCS-1611424 and ECCS-1933527), DTRA #BRCALL08-Per3-C-2-0006, and in part by SMART (2018-NE-2861), one of seven centers of nCORE, a Semiconductor Research Corporation program, sponsored by the National Institute of Standards and Technology (NIST). Work at LLNL was performed under the auspices of the U.S. DOE under Contract DE-AC52-07NA27344.

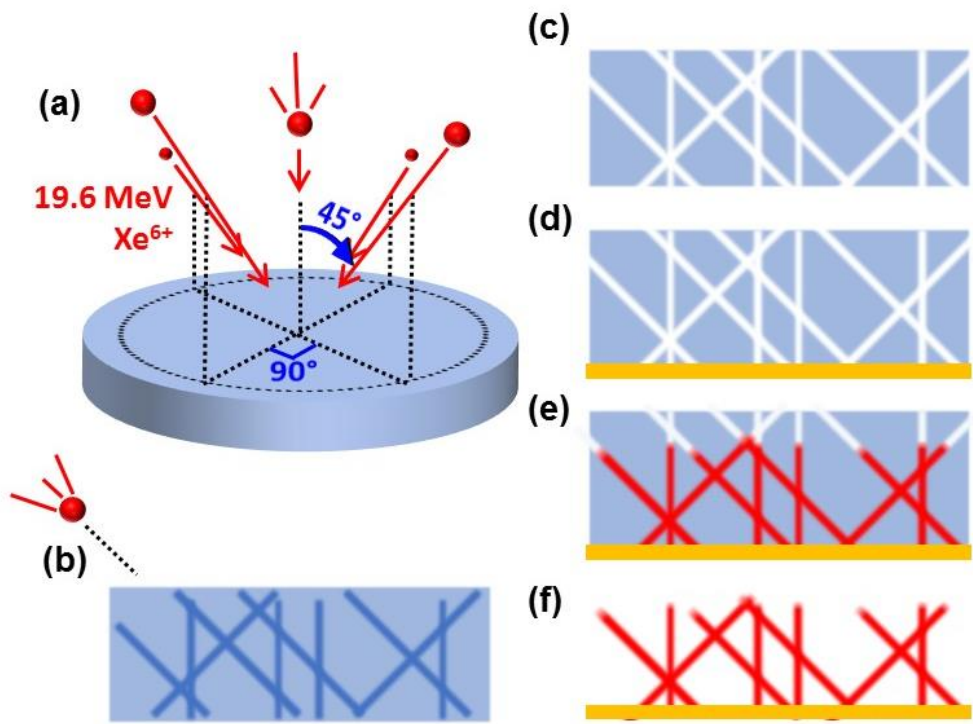

**Figure 1**: *Step-by-step production of the nanowire networks. Blue region represents the polycarbonate membrane. (a) Geometry of five ion-beam irradiations, (b) cross-section of polycarbonate showing latent tracks of ion damage (dark regions). (c) Etching of latent tracks to produce an interconnected nanoporous network. (d) Membrane with a deposited working electrode. (e) Electrodeposition of the metal network. (f) Removal of the polycarbonate membrane, leaving behind a freestanding lattice of wires atop the sputtered layer.*

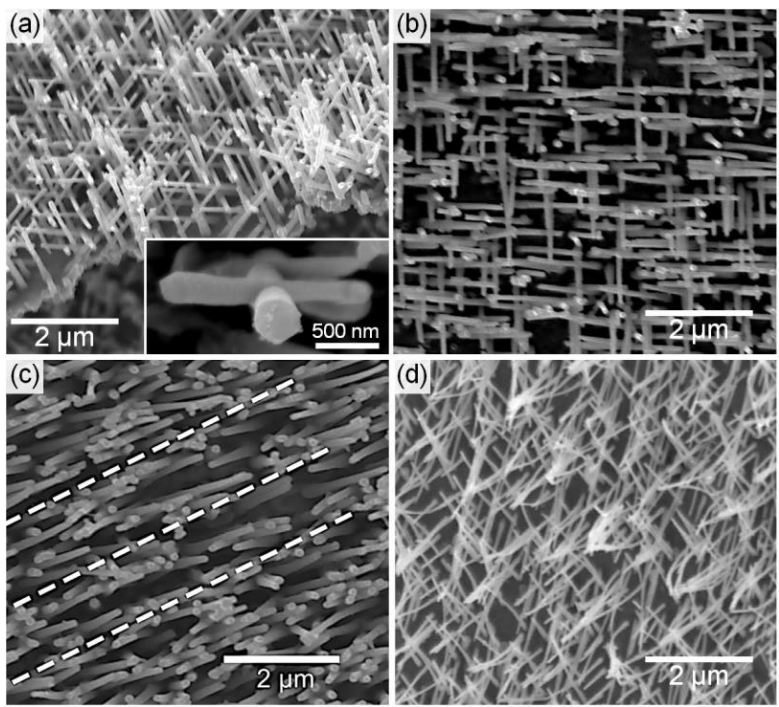

**Figure 2**: *(a) Side-view and (b) top-view SEM images of an interconnected copper network, tracked at five angles as described in the text, on a sputtered copper working electrode. Inset in (a) shows that at an intersection the wires are physically connected (from a different sample). Top-view SEM images of (c) a copper network and (d) a cobalt network, tracked at three angles. This produces intersections in planes that contain the three irradiated angles (which are highlighted with white dashed lines).*

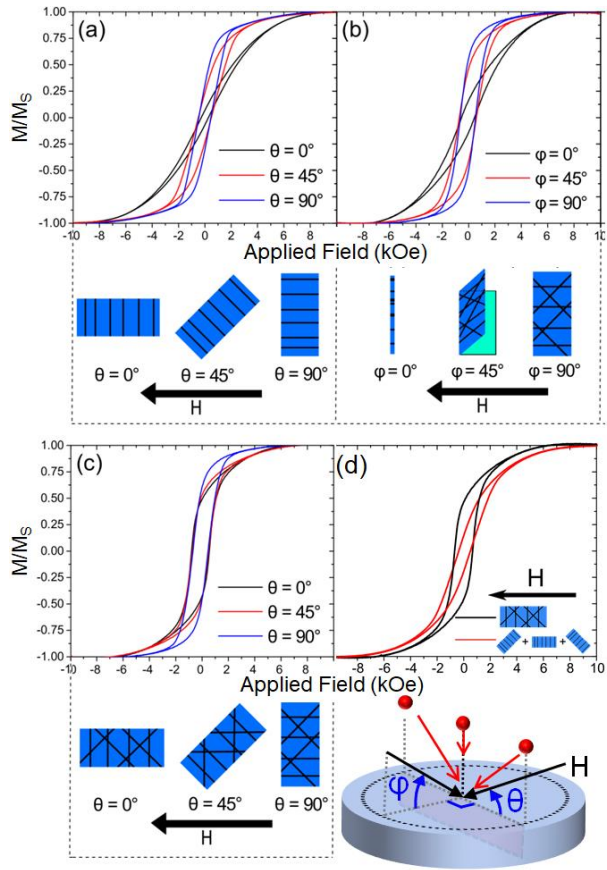

**Figure 3**: *Magnetic hysteresis loops measured by VSM. The measurement geometry is illustrated in the inset, where the three tracking angles are marked by red arrows, defining an irradiation plane; applied field H (black arrow) makes an angle ϑ in the irradiation plane with respect to the membrane, or an angle φ relative to the irradiation plane normal. Below each panel is a diagram illustrating the measurement geometry. The polycarbonate membrane is shown by the blue box, while the nanowires inside are shown by black lines. **(a)** Single angle irradiated sample measured at φ=90° and ϑ = 0°, 45°, and 90°, respectively. **(b,c)** Three-angle irradiated sample measured at different (b) φ axis and (c) ϑ axis (the applied field is always in the irradiation plane, with φ=90°). (d) Comparison of the hysteresis loop of intersecting wires measured at φ=90° and ϑ = 0° (black) vs. a reconstructed hysteresis loop (red). The latter is made by adding together non-intersecting wire loops oriented at the same angles to the applied field and weighted by the number of wires at those angles to mimic the intersecting sample.*

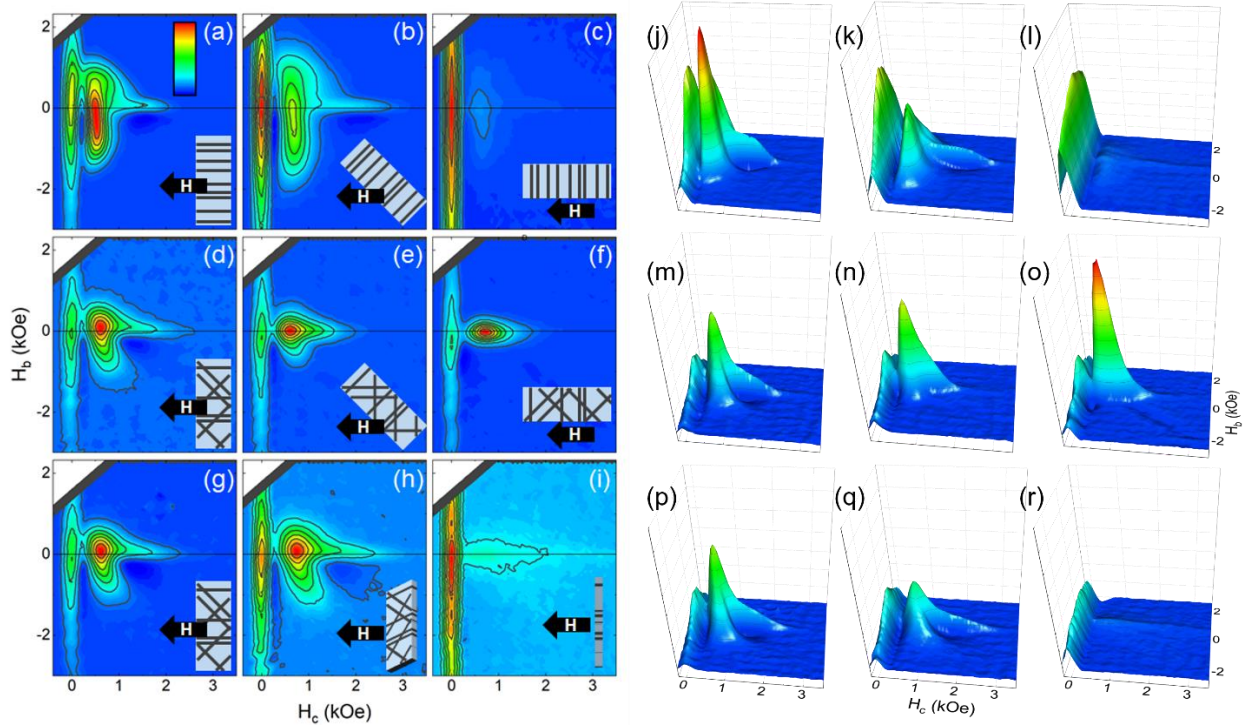

**Figure 4**: *FORC diagrams shown in 2D contour plots **(a-i)** and 3D distributions **(j-r)** for **(a-c, j-l)** a single-angle irradiated sample at ϑ=90°, 45° and 0°, respectively, and **(d-l, m-r)** three-angle irradiated sample, measured at φ=90° and ϑ = 90°, 45° and 0°, respectively **(d-f, m-o)**, and at ϑ = 90° and φ=90°, 45° and 0°, respectively **(g-i, p-r)**. The schematic diagrams in the lower right corner of panels **(a-i)** indicate the orientation of the wires (black lines) and membrane (blue box) relative to the applied field shown by the large black arrow. **(d)** and **(g)** represent the same geometry, so do **(m)** and **(p)**, but measured separately. The 2D FORC distribution in each panel **(a-i)** is scaled to its own maximum intensity to better illustrate the features, while the 3D FORC distributions in panels **(j-r)** are normalized to the integrated intensity for each sample.*

# 3D Nanomagnetism in Low Density Interconnected Nanowire Networks

Edward C. Burks,[1] Dustin A. Gilbert,[1,2,3] Peyton D. Murray,[1] Chad Flores,[1] Thomas E. Felter,[4]

Supakit Charnvanichborikarn,[5] Sergei O. Kucheyev,[5] Jeffrey D. Colvin,[5] Gen Yin,[6] and Kai Liu[1,6,*]

[1]*Physics Department, University of California, Davis, CA 95618*

[2]*Department of Materials Science and Engineering, University of Tennessee, Knoxville, TN 37996*

[3]*Department of Physics and Astronomy, University of Tennessee, Knoxville, TN 37996*

[4]*Sandia National Laboratories, Livermore, CA 94551*

[5]*Lawrence Livermore National Laboratory, Livermore, CA 94551*

[6]*Physics Department, Georgetown University, Washington, DC 20057*

**Number of intersections per unit area**

Given irradiation fluence *f* of each irradiated angle, the off-normal colatitude angle α, the thickness *t* of the template material, and the radius *r* of the resultant etched tracks, the total number of intersections per unit area *N* can be calculated as follows. This calculation assumes the pores have the same radius along their entire lengths (a good approximation for our samples), the off-normal angle was the same and in the same plane for the two off-normal irradiations (true for our samples) and it fails to account for a wire having multiple intersections (a decent approximation). This is done by calculating the intersection density of each of the four types of intersections: $N_1$, $N_2$, $N_3$ and $N_4$ as discussed below.

Cross-sectional illustrations of the four possible types of intersections are shown in Figure S1a, c, e and g, while Figure S1b, d, f and h show top views of their respective cases. For example, Figure S1a shows a cross-sectional view of a polycarbonate membrane (shown in blue) with two normally irradiated pores, one in red and one in purple, intersecting in a yellow region. Figure S1b depicts the same case viewed from the top, with the purple pore in an arbitrary location in the membrane and the yellow region around it showing the possible places the center of another pore could be located to cause an intersection. This area of possible intersection multiplied by irradiation fluence gives the average number of intersections that the particular purple pore will have with others, $n_1$:

$$n_1 = f(4\pi r^2) \tag{1}$$

But this is just the intersections caused by a single pore. To calculate the intersection density caused by this first type we must multiple by beam irradiation fluence again, giving:

$$N_1 = 4\pi f^2 r^2 \tag{2}$$

The intersection densities for the remaining types are calculated in the same way (see the calculated areas in Figure S1d, f and h):

$$N_2 = 2\pi f^2 r^2 (1 + \cos(\alpha)) + 4f^2 rt\sin(\alpha) \tag{3}$$

$$N_3 = 4\pi f^2 r^2 \cos(\alpha) \tag{4}$$

$$N_4 = 4\pi f^2 r^2 \cos(\alpha) + 4f^2 rt\sin(2\alpha) \tag{5}$$

Finally, to get the total intersection density $N$ we must add the totals from the four types and divide by two because we have counted each intersection twice using this method:

$$N = \frac{N_1 + N_2 + N_3 + N_4}{2} \tag{6}$$

For a total irradiation fluence of $2\times10^9$ tracks/cm² divided evenly across all three irradiations, a pore diameter of 75 nm, a membrane thickness of 3 μm, $\alpha = 45°$, and three angles, we get $N \approx 2\times10^9$ intersections/cm².

**Supporting Information**

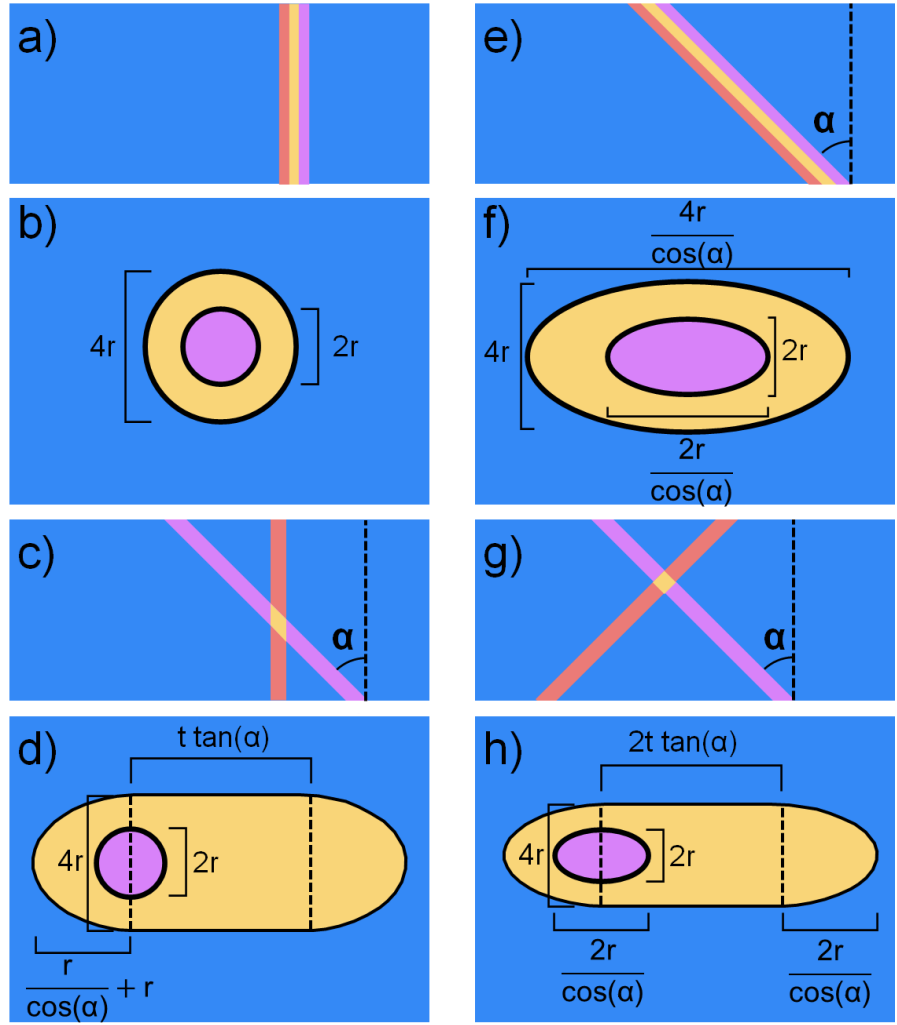

**Figure S1**. *Intersection probability is calculated by examining area of possible intersection in each of the four different intersection cases: two pores normal to the plane of the membrane (a,b), one pore normal and one at an angle α from the plane membrane (c,d), two pores at an angle α from the plane of the membrane (e,f) and one pore at an angle α and the other at an angle - α (g,h). In each case the first image shows a cross-section of the intersection and the second a top-down image for area calculation.*

3